# Deep Learning Improves Contrast in Low-Fluence Photoacoustic Imaging

ALI HARIRI,[1,†] KAMRAN ALIPOUR,[2,†] YASH MANTRI,[3] JURGEN P. SCHULZE,[2,4] AND JESSE V. JOKERST[1,5,6,*]

[1] *Department of NanoEngineering, University of California, San Diego, La Jolla, CA, 92093, USA*
[2] *Department of Computer Science, University of California, San Diego, La Jolla, CA, 92093, USA*
[3] *Department of BioEngineering, University of California, San Diego, La Jolla, CA, 92093, USA*
[4] *Qualcomm Institute, University of California, San Diego, La Jolla, CA, 92093, USA*
[5] *Department of Radiology, University of California, San Diego, La Jolla, CA, 92093, USA*
[6] *Material Science Program, University of California, San Diego, La Jolla, CA, 92093, USA*

*† These authors contributed equally to this paper.*

*\*jjokerst@ucsd.edu*

**Abstract:** Low fluence illumination sources can facilitate clinical transition of photoacoustic imaging because they are rugged, portable, affordable, and safe. However, these sources also decrease image quality due to their low fluence. Here, we propose a denoising method using a multi-level wavelet-convolutional neural network to map low fluence illumination source images to its corresponding high fluence excitation map. Quantitative and qualitative results show a significant potential to remove the background noise and preserve the structures of target. Substantial improvements up to 2.20, 2.25, and 4.3-fold for PSNR, SSIM, and CNR metrics were observed, respectively. We also observed enhanced contrast (up to 1.76-fold) in an in vivo application using our proposed methods. We suggest that this tool can improve the value of such sources in photoacoustic imaging.

## 1. Introduction

Photoacoustic imaging (PAI) combines the high-contrast of optical imaging and the high spatial resolution of ultrasound [1, 2]. Short optical pulses serve as an excitation source in PAI systems [3] to generate photoacoustic waves via thermoelastic expansion [4, 5]. Wideband ultrasonic transducers detect the propagated waves, and mathematical processing methods (reconstruction algorithms) can transform the detected signals into an image [6-10]. Over the last decade, investigators have demonstrated various applications of PAI in ophthalmology [11-13], oncology [14-16], dermatology [17-19], cardiology [20-22], etc.

PAI traditionally uses solid-state lasers as an excitation source because of their tunability, coherence, and high pulse energy. However these lasers difficult in clinical applications because they are bulky, expensive, unstable (in terms of power intensity fluctuation), and require frequent maintenance [23]. In contrast, pulse laser diodes (PLD) [24-26] and light emitting diodes (LED) [27-29] are a stable, affordable, and compact alternative light source. However, the output pulse energy of PLDs and LEDs is low—on the order of μJ/pulse and nJ/pulse versus mJ/pulse with lasers. Thus, the resulting photoacoustic data needs be averaged hundreds of time to cancel the noise and extract meaningful signal [27]. Unfortunately, performing many averages negatively affects the temporal resolution. Investigators have improved the signal-to-noise ratio (SNR) using classical signal processing methods such as empirical mode decomposition (EMD) [30, 31], wavelet-based methods [32, 33], Wiener deconvolution [34], principle component analysis (PCA) [35], and adaptive noise canceler (ANC) [36, 37]. However, these methods all require some prior information about the signal and noise properties, which is a significant limitation. Therefore, new tools to increase the SNR in low fluence PLD and LED PAI are needed.

Deep learning is rapidly expanding within various fields and improving performance in pattern recognition and machine learning applications. These relatively new techniques have vastly outperformed other classical methods in recent years. For example, computer vision has extensively utilized deep learning algorithms object detection, image classification, and image reconstruction [38, 39]. Convolutional neural networks (CNN) are among the most popular deep learning algorithms [40].

In medical imaging, previous studies focused on denoising CT-specific noise patterns. Kang et al. [41] utilized CNNs for wavelet transform coefficients with low-dose CT images. Chen et al. [42] used CNNs to directly map low-dose CT images to their normal-dose counterparts. Other methods altered the original CNN architecture to either preserve details in the image through *residual* blocks [43, 44] or *generator* CNNs to produce the restored image based on encoded features of low-dose images [45-47].

To capture more spatial context, previous approaches used pooling between convolution to reduce feature map resolution. However, pooling extends the receptive field and depth of their CNNs to drastically increase the computational costs in training and deploying such models [48]. Dilation [49] is another alternative to pooling but is limited by sparse sampling in the input layer, which can lead to gridding issues [50].

The concept of denoising in PA images is similar to low-dose CT yet the noise can have very different patterns; hence, the noise requires a different transfer method to be removed. Some of the earlier methods used short-lag spatial coherence [51, 52] or singular value decomposition (SVD) [53] to remove reflection artifacts from PA images. Some recent approaches utilize CNNs to identify point sources per image [54, 55] or use recurrent neural networks (RNN) to leverage temporal information in PA images to remove artifacts [56]. Antholzer et al. [57] adopted U-net architectures to reconstruct photoacoustic tomography (PAT). Anas et al. [58] utilized skip connections in dense convolutional networks to improve the quality of the PA images.

The potential of deep learning to enhance image quality motivates this work with a deep convolutional neural network. The goal is to map the low-fluence light photoacoustic images to corresponding high-fluence photoacoustic data. We demonstrated that deep learning can restore the images in low-fluence photoacoustic configuration with less computational cost than classical methods. Here, we first describe the proposed deep convolutional neural network and training details. We then demonstrate qualitative and quantitative phantom results. Finally, we show the capability our proposed model to image low concentrations of contrast agents *in vivo*.

## 2. Methods and Materials

We used a multi-level wavelet-CNN model with low receptive field, low computational cost, and high adaptivity for PA imaging in multi-frequency space [59]. The model is based on a U-Net architecture and consists of a contracting sub-network followed by an expanding subnetwork. The contracting subnetwork uses discrete inverse wavelet transform (DWT) instead of pooling operations. This substitution allows high-resolution restoration of image features through inverse wavelet transform (IWT) within the expanding subnetwork.

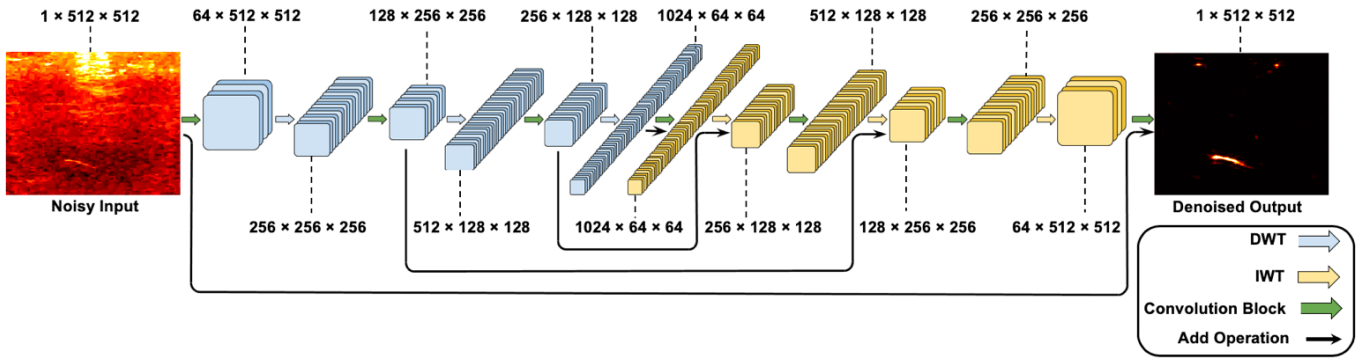

**Figure 1. MWCNN model architecture.** Contracting subnetwork features are extracted in wavelet space. In expanding the subnetwork, features expand into the image space while preserving high resolution details. This model takes a 512 x 512 noisy image as the input and transfers that to a 512 x 512 denoised output image. Add operations directly feed the contracting feature maps to expanding feature maps to preserve image details and avoid blur effects.

The model takes a 512 x 512 noisy image as the input and transfers that to a 512 x 512 denoised output image (**Figure 1**). The model processes the image in one channel as a heatmap. The input is a 2D cross-sectional framelet from a PA imagery set. The model attempts to reduce the noise in the image while preserving the signal. This model expands the feature dimension of the input from 1 to 1024 channels and then contracts the feature maps back into 1 channel as the output. The convolution blocks may contain multiple convolutional layers. Each convolution layer is followed by a ReLU activation function.

In the contracting subnetwork of the model, the image features go through multiple convolutions with intermittent DWT blocks. Our model uses a Haar wavelet transform based on the following orthogonal filters:

$$f_{LL} = \begin{bmatrix} 1 & 1 \\ 1 & 1 \end{bmatrix} \qquad f_{LH} = \begin{bmatrix} -1 & -1 \\ 1 & 1 \end{bmatrix} \qquad f_{HL} = \begin{bmatrix} -1 & 1 \\ -1 & 1 \end{bmatrix} \qquad f_{HH} = \begin{bmatrix} 1 & -1 \\ -1 & 1 \end{bmatrix}$$

The DWT blocks transform feature maps into four sub-bands. Due to the biorthogonal property of this operation, the original feature map can be accurately retrieved by an inverse Haar wavelet transform. The IWT blocks are then placed in between convolution blocks of the expanding subnetwork. For more details on the properties this transform, readers are referred to the original work [59].

Other CNN methods mostly use U-Net based architecture utilizing pooling operating in between convolutions—the average pooling in these systems can cause information loss in the feature maps. The MWCNN architecture benefits from DWT and IWT as a safe down-sampling and up-sampling processes where the feature maps can be transmitted with no information loss throughout the model. The objective of the training process is to optimize the model parameters $\Theta$ with the goal of minimizing the MSE loss function:

$$\mathcal{L}(\Theta) = \frac{1}{N} \sum_{i=1}^{N} \left|\left| y_i - F(x_i\,;\,\Theta) \right|\right|^2$$

The training set is $\left\{ (x_i, y_i) \right\}_{i=1}^{N}$. In this equation, $x_i$ is the low fluence (noisy) input image, $y_i$ is the corresponding high energy ground truth, and $F(x_i\,;\,\Theta)$ is the model output.

In PA imaging, the absolute magnitude of the signal and noise is dependent on multiple factors like light illumination, acoustic detection, and the experimental setup. Training a model based on the relative magnitude of the signal and noise might limit the model to specific types of samples and settings. Here, we focus the training on the shape features of the signal rather than the magnitude because such a model can be more generic and scalable. To minimize the model's reliance on the signal magnitude, we normalize the pixel values of the input between zero and one. In this setting, the model is inclined to distinguish noise from signal based on the *shape* features. The model is trained in a supervised manner to transform low energy inputs into outputs as close as possible to the ground truth frames.

*2.1 Photoacoustic Imaging System*

Two different commercially available pre-clinical photoacoustic imaging systems were used in this study. Both systems can physically translate the transducer to generate three-dimensional (3D) images. Model training used the Vevo LAZR (VisualSonic Inc.), which utilizes laser excitation integrated into a high frequency linear array transducer (LZ-201, Fc = 15 MHz) with optical fibers integrated to both sides of the transducer[60]. For optical excitation, this system uses a Q-switched Nd:YAG laser (4-6 ns pulse width) with a repetition rate of 20 Hz (frame rate of 6 Hz) followed by an optical parametric oscillator (tunable wavelength 680-970 nm). The laser fluence was 17.06 ± 0.82 mJ using a laser pyroelectric energy sensor (PE50BF-C, Ophir LLC, USA).

To modulate the intensity of the laser, we placed the sample in different concentrations of $TiO_2$ nanoparticles. These nanoparticles are well-known scatterers that decrease the fluence on the sample (when placed between the source and the sample). We measured the fluence with difference concentrations of nanoparticles with same energy sensor mentioned above.

Test data used a scanner lower fluence than the laser (LED excitation; AcousticX; CYBERDYNE Inc.) [27]. This system is equipped with a 128-element linear array ultrasound transducer with a central frequency of 10 MHz and a bandwidth of 80.9% fitted with two 690 nm LED arrays. The repetition rate of these LEDs is tunable between 1, 2, 3, and 4 K Hz. The pulse width can be changed from 50 ns to 150 ns with a 5-ns step size. The LED fluence at 1 K Hz and 2 K Hz was 20 and 40 µJ/pulse, respectively.

## 2.2 Image Evaluation Metrics

### 2.2.1 Peak Signal-to-Noise Ratio (PSNR)

We used the PSNR metric to evaluate the model in terms of noise cancelation. The PSNR is described in decibel (dB) and calculated based on square differences between the model output and ground truth images as:

$$PSNR = 20\log_{10}\left(\frac{I_{max}}{\sqrt{MSE}}\right),$$

where,

$$MSE = \frac{1}{MN} \sum_{m=0}^{M-1} \sum_{n=0}^{N-1} \left(I_{GT}(m,n) - I_{MWCNN}(m,n)\right)^2,$$

$I_{GT}$ and $I_{MWCNN}$ are the ground truth and model output images, respectively. Term $I_{max}$ represents the maximum possible value in the given images [56].

### 2.2.2 Structural Similarity Index Measurement (SSIM)

The SSIM evaluates image quality in terms of structural similarity; it is represented on a scale of 1. A higher SSIM shows better structural similarity of an output model image with the ground truth data.

$$SSIM = \frac{(2\mu_{GT}\mu_{MWCNN} + k_1)(2\sigma_{cov} + k_2)}{(\mu_{GT}^2 + \mu_{MWCNN}^2 + k_1)(\sigma_{GT}^2 + \sigma_{MWCNN}^2 + k_2)},$$

Here, $\mu_{GT}(\sigma_{GT})$ and $\mu_{MWCNN}(\sigma_{MWCNN})$ are the mean (variance) of the ground truth and MWCNN output images, respectively; $\sigma_{cov}$ shows the covariance between these two data. $k_1$ and $k_2$ are used to stabilize the division with a weak denominator [56].

### 2.2.3 Contrast-to-Noise Ratio (CNR)

CNR determines the image quality and is described in decibel (dB) via the following equation:

$$CNR = 20\log_{10}\frac{(\mu_{object} - \mu_{background})}{\sigma_{background}},$$

Here, $\mu_{object}$ and $\mu_{background}$ are mean of object and background noise, respectively; $\sigma_{background}$ represents standard deviation of background intensity in the image [61].

## 2.3 Training

The main aim of this study is to train the proposed convolutional network (MWCNN) to transform the low-fluence photoacoustic data into high-fluence images. We defined the high fluence images as the ground truth and then used TiO$_2$-based optical scatters to reduce the laser fluence (**Figure 2A**). The laser fluence was 17, 0.95, 0.25, 0.065, and 0.016 mJ/pulse at a wavelength of 850 nm with 0, 4, 6, 8, and 10 mg/ml of TiO$_2$, respectively **(Figure 7)**. Network training used a 3D pen print (Gincleey 3D Pen, AM3D Printers Inc.) to prepare a complicated 3-dimensional structure (2 cm x 2 cm x 3 cm) (**Figure 1B**). These structures were placed in an agarose phantom and scanned (30 mm, 270 frames) with all five optical filters on top of them (**Figure 1C**). We used 0 mg/ml which has the highest laser fluence (~17 mJ/pulse) as a ground truth in the proposed network. **Figure 1D** shows that the signal-to-noise ratio will decrease by decreasing the laser fluence (increasing the TiO$_2$ concentration).

In the training process, 85% of the frames (with fluence values of 17, 0.95, 0.25 mJ/pulse) were randomly selected as training set and the rest as test set. The training algorithm was implemented under the PyTorch platform using two NVIDIA GeForce GTX 1080 Ti GPUs. We used ADAM optimizer for our training algorithm with an initial learning rate of $1.024 \times 10^{-4}$. The training process completed 256 epochs in one day.

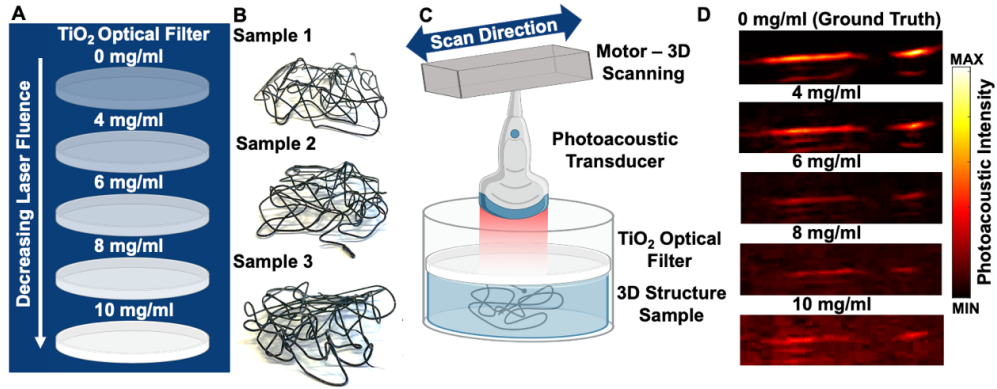

**Figure 2. Experimental training setup. A)** TiO$_2$-based optical scattering gels reduce the laser fluence. The laser fluence at wavelength of 850 nm was 17, 0.95, 0.25, 0.065, and 0.016 mJ/pulse after using 0, 4, 6, 8, and 10 mg/ml of TiO$_2$, respectively. **B)** Three different complicated 3D structures were made using a 3D pen print to collect training data. **C)** Imaging setup—the 3D structures are placed in the agarose phantom with a TiO$_2$-based optical scatter on top. We scanned the entire structure for each sample and acquired 270 frames; 850 nm was the illumination wavelength. **D)** B-mode photoacoustic images with different optical scatterer concentrations and thus laser fluence values show reduced SNR with increasing TiO$_2$ concentration (decreased laser fluence). Scale bars represent 1 cm.

## 2.4 Testing

### 2.4.1 Low Fluence Laser Source

To test the trained MWCNN model, we printed term "UCSD" on a transparent film in blank ink and placed it beneath the agarose hydrogel. The black ink is strongly absorbing and will produce photoacoustic signal. The TiO$_2$ optical scatters with the same concentrations as in the training section were used to test the model under different laser fluence values. We collected 270 frames, and each frame was individually used as an input in the MWCNN model. Testing was done at 0.065 and 0.016 mJ/pulse fluences to evaluate model scalability in illuminations lower than the training domain (0.95 and 0.25 mJ/pulse). All output frames were placed next to each other to generate the 3D volumetric data. Importantly, the trained MWCNN model was totally blind to this new data set. We measured the PSNR and SSIM metrics on each letter: U, C, S, and D. We performed t-test statistical analysis, and $p$ values < 0.05 were considered to be significantly different.

*2.4.2 LED-Based Light Source*

We also tested the MWCNN model with the LED-based photoacoustic imaging system but without any nanoparticle gel scatterers (LED system has inherently low fluence). We again printed "UCSD" and placed it beneath a transparent agarose hydrogel. The LED was operated at 1 K Hz and 2 K Hz for 40 and 80 µJ/pulse on the sample. We used all 180 frames as the input for the model. We placed all frames after each other to create a 3D volumetric image. The ground truth data were collected by operating the LED source at 4 K Hz (160 µJ of fluence). We used 20 rounds of averaging for each frame. PSNR and SSIM metrics were calculated for each letter.

We also evaluated the model with an LED-based system and a different configuration. We placed pencil lead (0.5 mm HB, Newell Rubbermaid, Inc., IL, USA) at depths of 2.5, 7.5, 12.5, 17.5, and 22.5 mm in 2% intralipid (20%, emulsion, Sigma-Aldrich Co, MO, USA) mixed with agarose. We used intralipid to mimic biological tissue. We collected a single frame with the LED system at 1 K Hz and 2 K Hz to have 40 and 80 µJ/pulse on the surface of the sample. To measure the CNR for both input (Noisy) and output (MWCNN model) images, $\mu_{object}$ and $\mu_{background}$ were defined as the average (5 different areas) of the mean photoacoustic intensity on the pencil lead (ROI of 1 X 1 mm²) and the background area (ROI of 3 X 3mm²), respectively. Term $\sigma_{background}$ is the average of all five standard deviations of background intensity.

*2.4.3 In vivo Performance*

We also evaluated our trained model in its ability to enhance the contrast agent detection in vivo. Here, the murine tissue reduces the fluence. We purchased nine nude mice (8-10 weeks) from the University of California San Diego Animal Care and Use Program (ACP). All animal experiments were performed in accordance with NIH guidelines and approved by the Institutional Animal Care and Use Committee (IACUC) at the University of California, San Diego. The mice were anesthetized with 2.5% isofluorane in oxygen at 1.5 L/min. Methylene blue (MB) (Fisher Science Education Inc., PA, USA) was purchased and dissolved in distilled water. MB concentrations of 0.01, 0.05, 0.1, 1, and 5 mM were injected intramuscularly in a murine model (n=3). The Vevo LAZR (VisualSonic Inc.) system was used for this in vivo experiment. We monitored the injection procedure using both ultrasound and photoacoustic images. The location of injected MB was confirmed using photoacoustic spectral data. We measured the CNR for both input (Noisy) and output (MWCNN model) images. For this calculation, $\mu_{object}$ and $\mu_{background}$ were defined as tahe verage (5 different areas) of mean values of photoacoustic intensity at the injected area ( ROI of 1 X 1 mm²) and around the injected area (ROI of 3 X 3mm²), respectively. Term $\sigma_{background}$ is average of all five standard deviations of background intensity.

### 3. Results

*3.1. Low Fluence Laser Source*

Laser fluence decreases when passing through a scattering media such as biological tissue. The photoacoustic signal intensity is proportional to the laser fluence, and thus the quality of acquired images will be affected. We intentionally decreased the laser fluence and improved the acquired images using MWCNN.

**Figure 3A** illustrates the ground truth 3D image of "UCSD" using the full laser fluence (17 mJ/pulse) from the Vevo LAZR system. **Figure 3B, C** shows the SSIM and PSNR vs laser fluence for both noisy (input) and MWCNN model (output) images. This model improved the SSIM with a factor of 1.45, 1.5, and 1.62 for laser fluence values of 0.95, 0.25, and 0.065 mJ/pulse, respectively. The PSNR was enhanced by 2.25-, 1.84-, and 1.42-fold for laser fluence values of 0.95, 0.25, and 0.065 mJ/pulse, respectively. The trained MWCNN model cannot significantly improve SSIM and PSNR ($p > 0.05$) at 0.016 mJ/pulse. **Figure 3D, E, F,** and **G** shows the noisy (input) 3D image of the UCSD object captured with laser fluence values of 0.95, 0.25, 0.065, and 0.016 mJ/pulse, respectively. The output of MWCNN for laser fluence of 0.95, 0.25, 0.065, 0.016 mJ is represented in **Figure 3H, I, J,** and **K**, respectively.

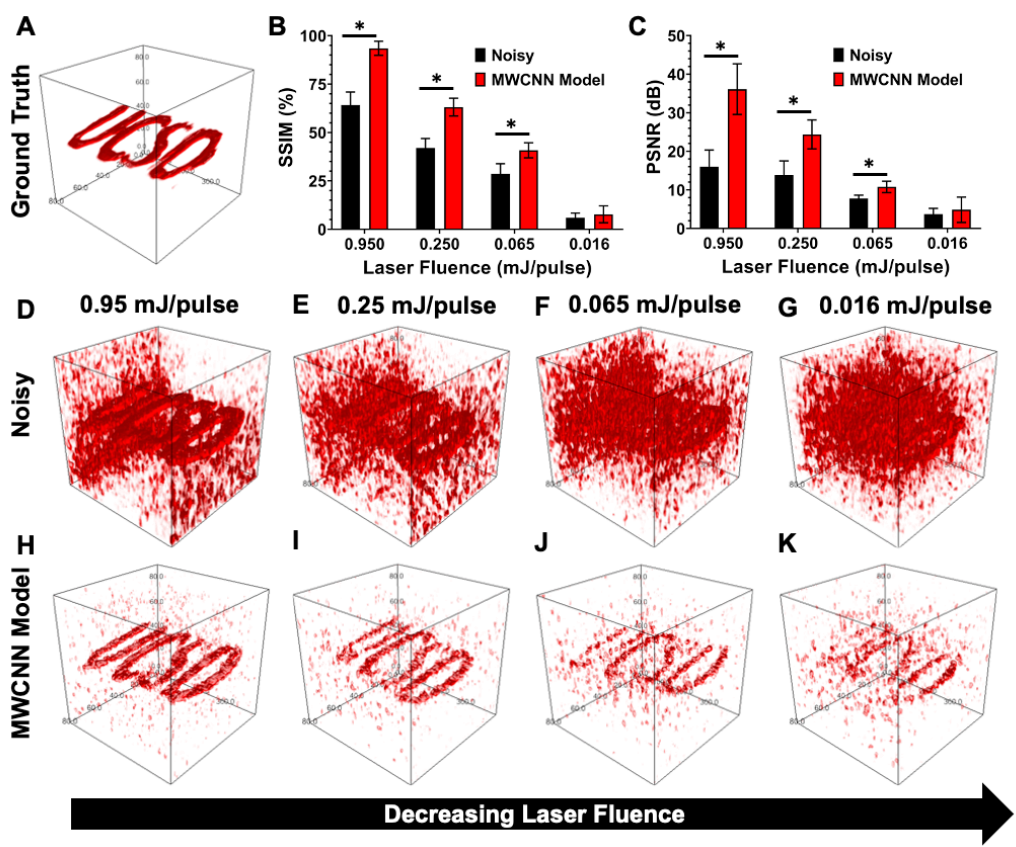

**Figure 3. Low fluence laser source evaluation. A)** Ground truth 3D image of UCSD sample with full laser fluence of 17 mJ/pulse. We used this image as a reference for measuring image quality metrics. **B)** SSIM of noisy (input) and MWCNN data vs laser fluence. The results show that the SSIM is significantly improved by 1.45, 1.5, and 1.62 at laser fluence values of 0.95, 0.25, 0.065 mJ/pulse, respectively. The model failed to improve the structural similarity at fluence of 0.016 mJ/pulse. **C)** PSNR of both noisy and MWCNN data vs laser fluence—the PNSR is significantly improved with a factor of 2.25, 1.84, and 1.42 for 0.95, 0.25, 0.065 mJ/pulse, respectively. However, the MWCNN cannot significantly improve the image quality with a laser fluence of 0.016 mJ/pulse. In both B and C, the error bars represent the standard deviation of SSIM and PSNR among the four letters in "UCSD". * indicates $p < 0.05$. **D, E, F, and G)** Noisy (input) images with 0.95, 0.25, 0.065, and 0.016 mJ/pulse laser fluence, respectively. **H, I, J,** and **K)** MWCNN model (output) images with 0.95, 0.25, 0.065, and 0.016 mJ/pulse laser fluence, respectively.

## 3.2. LED Source

We next examined our MWCNN model with LEDs as a low fluence source. **Figure 4A** shows the ground truth 3D photoacoustic image using the LED-based photoacoustic imaging system. After collecting all the noisy data with 40 and 80 µJ/pulse as LED fluence on the sample, we noted an improvement in SSIM by a factor of 2.2 and 2.5 for 40 and 80 µJ/pulse, respectively **(Figure 4B)**. **Figure 4C** demonstrates a 2.1- and 1.9-fold increase in PNSR on MWCNN model (output) versus noisy (input) for both 40 and 80 µJ/pulse. **Figure 4D** and **F** shows the noisy 3D photoacoustic image using the LED-based imaging system with fluence values of 80 and 40 µJ µJ/pulse. **Figure 4E** and **G** are MWCNN 3D results with 80 and 40 µJ µJ/pulse, respectively.

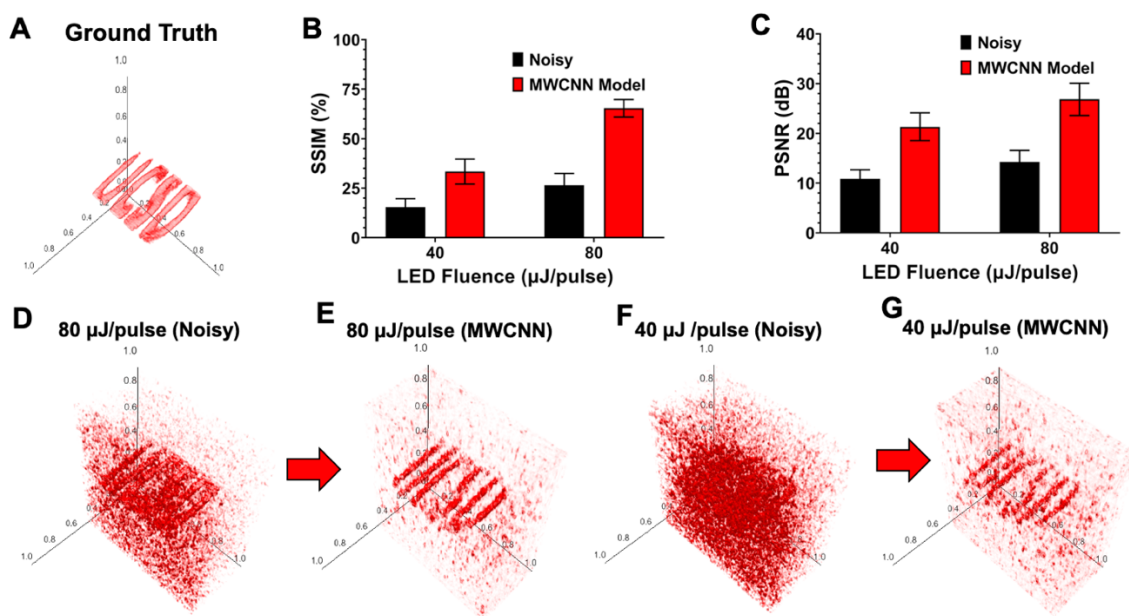

**Figure 4. LED light source evaluation. A)** Ground truth 3D image of UCSD word using the LED-based photoacoustic imaging system. The ground truth data were collected by operating the LED source at 4 K Hz with a fluence of 160 µJ and 20 rounds of averaging for each frame. **B)** SSIM results of both noisy (input) and MWCNN model (output) in two different LED fluences of 40 and 80 µJ/pulse. An improvement of 2.2- and 2.5-fold is observed for 40 and 80 µJ/pulse, respectively. **C)** PSNR of noisy (input) and MWCNN model (output) at 40 and 80 µJ/pulse. MWCNN improved the PNSR by 2.1 and 1.9 for 40 and 80 µJ/pulse, respectively. In both B and C, the error bars represent the standard deviation of SSIM and PSNR among the four letters in "UCSD". **D)** The 3D noisy (input) photoacoustic image used 80 µJ/pulse. **E)** 3D MWCNN mode (output) photoacoustic image using 40 µJ/pulse. **F)** 3D Noisy (input) photoacoustic image with 40 µJ/pulse. **G)** 3D MWCNN mode (output) photoacoustic image with 40 µJ/pulse.

We also evaluated the performance of the MWCNN model in penetration depth phantoms using the LED system and scattering media. **Figure 5A** and **B** shows noisy images for LED fluence at 40 and 80 µJ/pulse. We observed a significant CNR improvement both qualitatively (**Figure 5C** and **D**) and quantitatively (**Figure 5E**). We measured an average of 4.3- and 4.1-fold enhancement in MWCNN model versus noisy images at different depths for 40 and 80 µJ/pulse. **Figure 5E** shows that the MWCNN also enhanced the linearity of CNR vs depth from $R^2$ =0.84 and 0.85 to $R^2$ =0.97 and 0.95 for both LED fluence values.

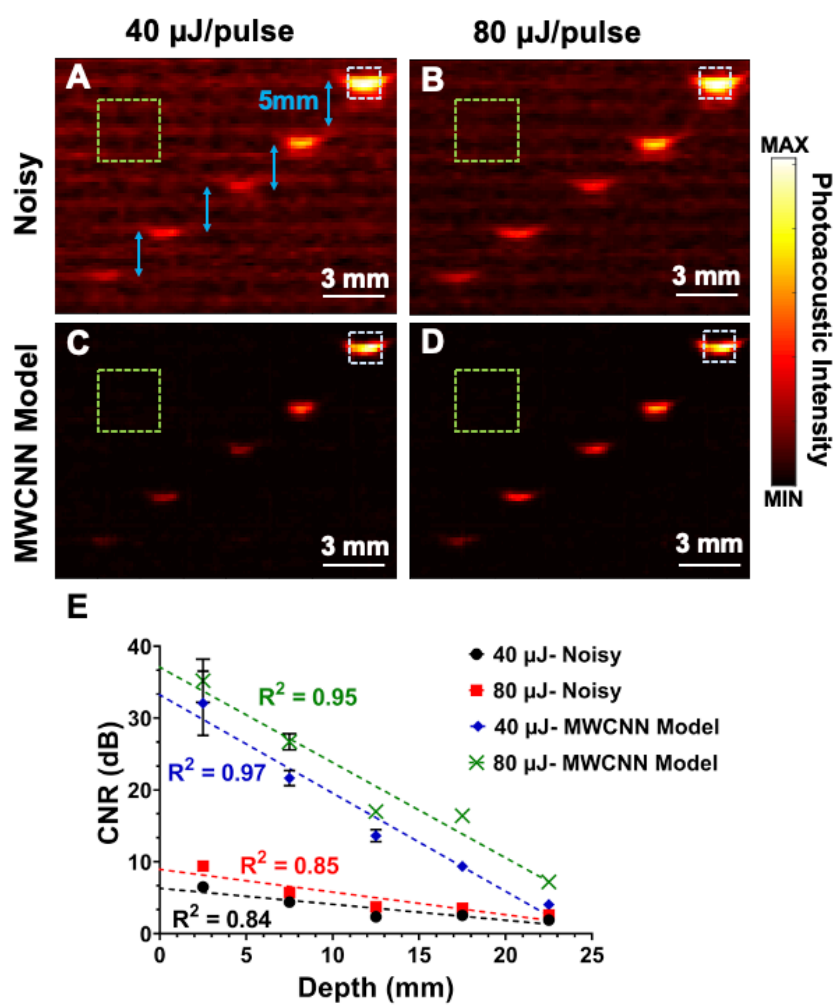

**Figure 5. Penetration depth evaluation using an LED. A)** B-mode noisy (input) photoacoustic image using LED at a fluence of 40 µJ/pulse. Pencil leads were placed at 2.5, 7.5, 12.5, 17.5, and 22.5 mm in 2% intralipid. **B)** B-mode noisy (input) photoacoustic images at a fluence of 80 µJ/pulse with similar experimental setup as described in A. **C, D)** B-mode MWCNN model (output) photoacoustic image for 40 and 80 µJ/pulse. **E)** CNR versus depth for 40 and 80 µJ/pulse in both noisy and MWCNN model. Dotted green and white rectangles represent the ROI used to measure mean values and standard deviations of background (ROI size:3 x 3 mm²) and object (ROI size:1 x 1 mm²). We observed an average of 4.3- and 4.1-fold enhancement in the MWCNN model versus noisy data at different depths for both LED values.

### 3.3. In vivo Performance

Image enhancement methods become more valid when validated *in vivo*. The detection of exogenous contrast agents using photoacoustic imaging technique can be a challenge due to low fluence due to scattering by biological tissue. Here, we injected different concentrations of MB intramuscularly in mice (**Figure 6A**) and analyzed the acquired images using the MWCNN model and compared the CNR with and without the model. **Figure 6B** represents the CNR for both noisy (input) and MWCNN model (output) for all 5 different concentrations. We observed a significant improvement between noisy and model output. There was an improvement of 1.55, 1.76, 1.62, and 1.48 in CNR for 0.05, 0.1, 1.0, and 5.0 mM, respectively. The MWCNN failed to improve the CNR for 0.01 mM. The signal intensity of 0.01 mM MB was so low that the model considered it to be noise.

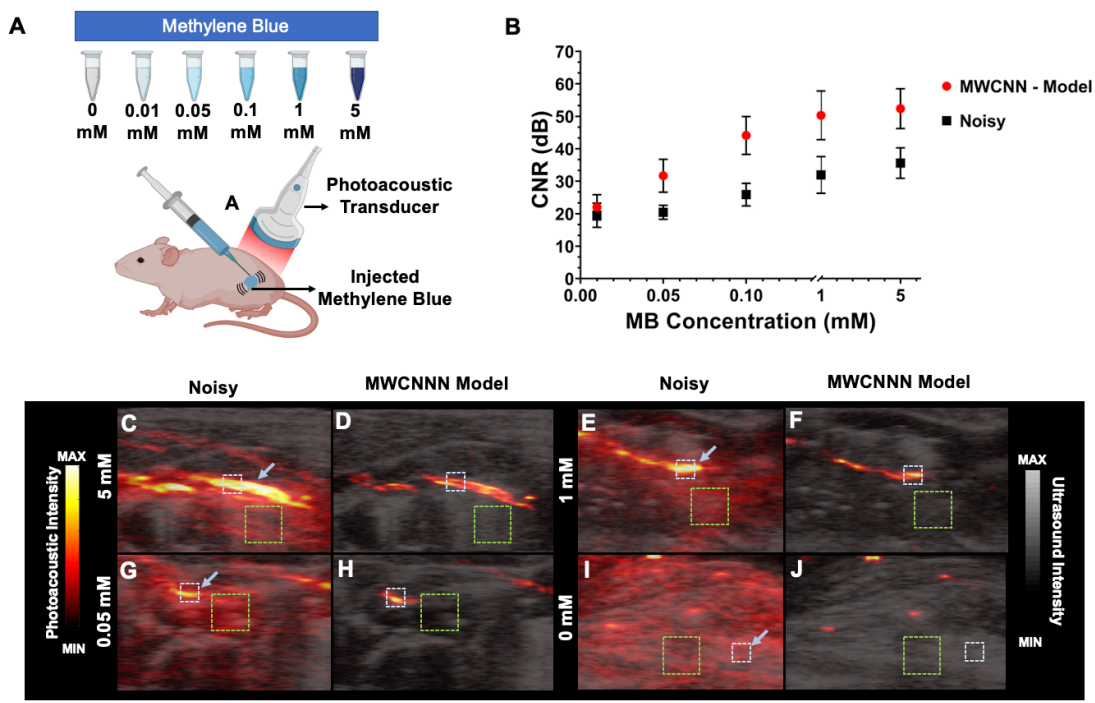

**Figure 6. In vivo evaluation of MWCNN model. A)** Experimental schematic for in vivo evaluation of MWCNN model. Five different concentrations of MB (0.01, 0.05, 0.1, 1, and 5 mM) were injected intramuscularly. **B)** CNR versus injected MB concentrations for both noisy and MWCNN model. We noted 1.55-, 1.76-, 1.62-, and 1.48-fold improvement of CNR for 0.05, 0.1, 1.0, and 5.0 mM, respectively. Error bars represent the CNR among three different animals. For CNR calculation, $\mu_{object}$ and $\mu_{background}$ were defined as the average of five different areas of mean values of photoacoustic intensity at the injected area (ROI of 1 X 1 mm²) and around the injected area (ROI of 3 X 3mm²), respectively. Term $\sigma_{background}$ is the average of all five standard deviations of background intensity. Panels **C, E, G,** and **I)** are B-mode noisy photoacoustic images for 5.0, 1.0, 0.05, 0 mM, respectively. These images are overlaid on ultrasound data. **D, F, H,** and **J)** B-mode MWCNN photoacoustic images for 5, 1, 0.05, 0 mM, respectively. Dotted green and white rectangles represent the used ROIs for background and object. Blue arrows show the MB injection area.

## 4. Discussions and Conclusions

CNNs have been widely utilized in computer vision, image processing, and medical imaging. However, deep learning has utility beyond image segmentation, object detection, and object tracking. Here, we proposed a deep learning model that can learn to restore PA images at different low fluence configurations and samples. To ensure the scalability of our solution, we built our model based on a limited training process and evaluated it with different illumination sources on other sample types and materials.

We observed quantitative and qualitative enhancement results. The proposed model was completely blind to our test data. We could achieve up to 1.62- and 2.2-fold improvement in SSIM (**Figure 3B** and **4B**) for low fluence laser source and LED, respectively. The model improved the PSNR by a factor of up to 2.25 and 2.1 for low fluence laser and LED, respectively. A higher number of training datasets can lead to improvement factors (SSIM, PSNR) that will be significantly higher. PNSR and SSIM calculations require a ground truth image. However, having this data is not feasible in most cases. To show that our proposed method can enhance other image quality metrics, we used the CNR to evaluate the penetration depth and in vivo data. The ground truth is not required with this metric, and the CNR will be measured using just a single frame. We showed that the MWCNN can improve the contrast as well (**Figure 5E, 6B**). Finally, we showed that this contrast improvement has value in vivo with contrast improved up to 1.76-fold.

Like other deep learning methods, our solution gains most of its computational cost at the training stage. The training cost can scale up as the training set grows. At runtime, the model can process each frame at 0.8 seconds, which is relatively faster than classical methods like BM3D (3.33 seconds). It is also similar to DL methods like low-dose CT CNN (2.05 seconds) [42].

Our training included a small set of frames from a laser source within a specific range of illumination fluencies. Such a small training set can facilitate a model that trains fast for practical solutions. On the other hand, normalizing the training data made the model independent of signal magnitude in the input. This independence guided the model to generically learn important spatial features of samples in PA images beyond the settings and configurations.

The tests introduced lower fluencies of illumination from different sources. The observations suggest that the model learned features to distinguish signal from noise regardless of the input image quality. Comparable results between laser and LED based inputs also suggests the utility of our solution among various imaging systems.

The next step of this work will focus on training and testing processes on *in vivo* samples including actual blood vessels and other exogenous contrast agents. We will also expand the model from a 2D framework to 3D data. In that regard, we can train models based on a stack of PA images to potentially improve the consistency of results along the axis and reduce the noise in 3D results as well as cross-sectional images. Increasing the dimensions of the model will inherently increase the amount of training data required to develop the models but may facilitate even more advanced in vivo imaging.

## 5. Acknowledgements

JVJ acknowledges funding from the National Institutes of Health under grants R21 AG065776, R21 DE029025, and DP2 HL137187. We acknowledge NSF funding under grants 1842387 and 1937674. Infrastructure for this work was supported under NIH grant OD S10021821. Figures 2A, 2C, and 6A were created with BioRender.com.

## 6. Disclosures

The authors declare that there are no conflicts of interest related to this article.

## 7. Appendix

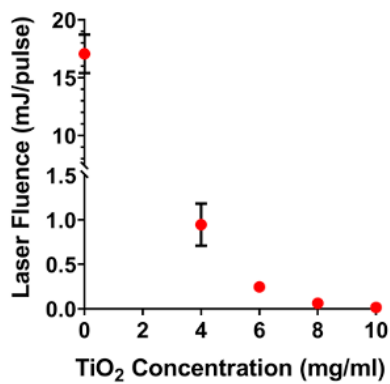

Figure 7. laser fluence vs TiO$_2$-based optical scattering concentrations